\documentclass[sigconf, nonacm]{aamas}

\usepackage{balance} 
\usepackage{multirow}

\setcopyright{ifaamas}
\acmConference[AAMAS '24]{Proc.\@ of the 23rd International Conference
on Autonomous Agents and Multiagent Systems (AAMAS 2024)}{May 6 -- 10, 2024}
{Auckland, New Zealand}{N.~Alechina, V.~Dignum, M.~Dastani, J.S.~Sichman (eds.)}
\copyrightyear{2024}
\acmYear{2024}
\acmDOI{}
\acmPrice{}
\acmISBN{}

\acmSubmissionID{149}

\title[AAMAS-2024 Formatting Instructions]
{Deconstructing Cooperation and Ostracism\\via Multi-Agent Reinforcement Learning}

\author{Atsushi Ueshima}
\affiliation{
  \institution{Tohoku University}
  \city{Sendai}
  \country{Japan}}
\email{ueshima73@gmail.com}

\author{Shayegan Omidshafiei}
\affiliation{
  \institution{Google Research}
  \city{Cambridge, MA}
  \country{USA}}
\email{somidshafiei@google.com}

\author{Hirokazu Shirado}
\affiliation{
  \institution{Carnegie Mellon University}
  \city{Pittsburgh, PA}
  \country{USA}}
\email{shirado@cmu.edu}

\begin{abstract}
Cooperation is challenging in biological systems, human societies, and multi-agent systems in general. While a group can benefit when everyone cooperates, it is tempting for each agent to act selfishly instead. Prior human studies show that people can overcome such social dilemmas while choosing interaction partners, i.e., strategic network rewiring. However, little is known about how agents, including humans, can learn about cooperation from strategic rewiring and vice versa. Here, we perform multi-agent reinforcement learning simulations in which two agents play the Prisoner's Dilemma game iteratively. Each agent has two policies: one controls whether to cooperate or defect; the other controls whether to rewire connections with another agent. This setting enables us to disentangle complex causal dynamics between cooperation and network rewiring. We find that network rewiring facilitates mutual cooperation even when one agent always offers cooperation, which is vulnerable to free-riding. We then confirm that the network-rewiring effect is exerted through agents' learning of ostracism, that is, connecting to cooperators and disconnecting from defectors. However, we also find that ostracism alone is not sufficient to make cooperation emerge. Instead, ostracism emerges from the learning of cooperation, and existing cooperation is subsequently \emph{reinforced} due to the presence of ostracism. Our findings provide insights into the conditions and mechanisms necessary for the emergence of cooperation with network rewiring.
\end{abstract}

\keywords{Cooperation, Ostracism, Network rewiring, Multi-agent reinforcement learning}

\newcommand{\BibTeX}{\rm B\kern-.05em{\sc i\kern-.025em b}\kern-.08em\TeX}

\begin{document}


\pagestyle{fancy}
\fancyhead{}


\maketitle 


\section{Introduction}

Understanding how cooperation arises and is sustained is crucial to comprehending biological systems, human societies, and multi-agent systems in general~\cite{Axelrod1981-xy, Nowak2006-bl, Olson1965-bc}. Cooperation can be difficult to achieve when it creates a social dilemma~\cite{Dawes1980-il}. In such situations, groups can benefit when everyone cooperates, but it is tempting for each agent to act selfishly instead.  Hence, encouraging individuals to develop and sustain cooperation presents significant challenges~\cite{Hardin1968-qq, Ostrom2000-qw, Leibo2017-dq}. How can individual agents learn to overcome short-sighted selfishness and achieve mutual cooperation for the greater good?

One possible way to address social dilemmas is network rewiring~\cite{Perc2010-ac, Rand2011-hv, Li2020-yl, Akcay2018-ym}. Individuals might not necessarily be bound to interact with their counterparts every time; instead, they might have some agency to establish and sever connections and choose whether to interact with them next time. Network rewiring offers an additional means of responding to the past actions of others. Not only can agents reciprocate by strategically changing their own cooperation behavior, but they can also alter their network connections, engaging in "tie reciprocity" or "ostracism"~\cite{Hirshleifer1989-vb, Shirado2013-nl, Feinberg2014-jr}. This paper defines ostracism as a specific network-rewiring policy in which agents connect to cooperators and disconnect from defectors. Prior simulations and human-subject experiments demonstrate that cooperation emerges with network rewiring, and most individuals exhibit such strategic network rewiring during the process~\cite{McNamara2008-io, Rand2011-hv, Shirado2013-nl, Melamed2018-bi, Fehl2011-kq, Fu2008-yk}.

However, little is known about how agents spontaneously learn about cooperative policies from network rewiring and vice versa. Previous methodology has limitations in disentangling complex causal dynamics between cooperation and network rewiring. Human-subject experiments, for instance, do not give participants sufficient time and iterations to develop policies through interaction with others~\cite{Rand2011-hv, Shirado2013-nl, Mao2017-yg}. Similarly, computational simulations often use a predetermined set of behavioral policies for agents in cooperation and network-rewiring decisions~\cite{Nowak2006-bl, Perc2010-ac, Hilbe2013-wr, Nowak1993-jy}, constraining their ability to examine policy generation. Recent works using multi-agent reinforcement learning (MARL) allow agents to generate behavioral policies in cooperative domains~\cite{Leibo2017-dq, De_Cote2006-kp, Koster2022-jn, McKee2023-hk}. However, they seek to design interactions and optimization mechanisms at the group level to explicitly facilitate cooperation, rather than investigate \emph{why} cooperation emerges in certain conditions. 

Here, we investigate how cooperation and ostracism are learned through the Iterated Prisoner's Dilemma (PD)~\cite{Axelrod1981-xy} in a two-player MARL environment. Two agents play the game using two policies: one policy controls whether to cooperate or defect with the other agent in a PD payoff structure (i.e., ~\textit{interaction policy}), while the other policy controls whether to rewire (establish or sever) connections with the other agent (i.e., ~\textit{network-rewiring policy}). Each agent has two distinct neural networks for each policy that are updated through dyadic interactions. In contrast to prior works, extrinsic institutions favoring cooperation are not present in our setting. We manipulate two key dimensions in this study: 1) the frequency of network-rewiring opportunities and 2) the interaction or network-rewiring policy for one agent. By locking one agent's policy, we examine how network rewiring encourages or discourages agents from learning about cooperation and ostracism from their counterparts. We evaluate the cooperation level of the two agents across the treatments as a consequence of reinforcement learning. 

Our findings show that network rewiring facilitates mutual cooperation even when one agent always offers cooperation, which is vulnerable to free-riding~\cite{Nowak2006-bl}. We also confirm that the network-rewiring effect is exerted through agents' learning of ostracism. However, ostracism alone is not sufficient to make cooperation emerge. Instead, ostracism emerges from the learning of cooperation, and existing cooperation is subsequently \emph{reinforced} due to the presence of ostracism. Our study provides valuable insights into the conditions and mechanisms necessary for the emergence of cooperation with network rewiring.

Our work makes the following contributions: 1) we develop a MARL framework to explore the coevolution of cooperation and population structure, 2) we uncover the complex cause-effect relationships involved in the dual learning of cooperation and network rewiring, and 3) we demonstrate the potential application of MARL in addressing important questions in system biology, social sciences, and humanities.


\section{Related work}
\paragraph{\textbf{Evolutionary games.}} Evolutionary game theory has been used to analyze cooperation difficulties by extending traditional game theory to long-term, multi-player scenarios~\cite{Nowak2006-bl,tuyls2005evolutionary}. For instance, Nowak and May utilized the theoretical framework to demonstrate that the evolution of behavioral policies alone is insufficient to alter social dynamics that attract individuals to defection~\cite{Nowak1993-jy}. Their research and subsequent studies suggest that certain interaction structures, such as network topology~\cite{Ohtsuki2006-ye} and heterogeneity~\cite{Santos2008-qf}, could help cooperators spread over an entire population while resisting defection. Several other works have applied coevolution rules to evolutionary games, indicating that cooperation can emerge as interaction networks evolve~\cite{Perc2010-ac, Fu2008-yk}.

Evolutionary games originally focus on intergenerational evolution, where agents with higher fitness reproduce their strategies more frequently in a population~\cite{Nowak2006-bl}. These simulations prepare for two or more strategies (or fixed policies) to interact with each other and update the fraction of each strategy based on their earnings or fitness. However, this assumption and setting do not allow us to scrutinize the emergence of cooperation in the real world through social learning. In real-world interindividual interactions, agents cannot directly observe or \emph{exactly} mimic others' policies. We bypass this issue in our study by applying MARL to network-rewiring settings, allowing agents to learn both policies to cooperate and to change interaction connections by observing an environment (including the other agent's behaviors) and obtaining rewards.

\paragraph{\textbf{Human studies.}} Evolutionary game theory has been tested through human-subject experiments to study human cooperation~\cite{Rand2013-jj}. By allowing human subjects to play a model game accordingly, researchers have examined what can make people cooperate. For instance, Rand, Arbesman, and Christakis conducted a lab experiment where human participants played a multi-player Prisoner’s Dilemma game repeatedly, changing their interaction partners in a sizable group~\cite{Rand2011-hv}. The results showed that cooperation stabilized at a high level through network rewiring while it decayed over time when their partners were shuffled or fixed. Shirado et al. clarified that people establish connections with cooperators and sever them from defectors, and the specific network-rewiring policy helps cooperators cluster and benefit each other ~\cite{Shirado2013-nl}.

Human-subject experiments have also revealed population-wise differences in cooperation~\cite{Henrich2001-ph}, wherein some populations admit egocentrism, while others show altruism in cooperation games.
These findings suggest that behavioral experiments allow us to study what policies humans have, but not how humans develop such policies through interaction. The experimental sessions, which usually last for 30-45 minutes (and for 20 days at longest~\cite{Mao2017-yg}), are simply not enough to examine the longer-term processes of developing and establishing individual policies.

\paragraph{\textbf{Multi-agent reinforcement learning.}} MARL has been used to overcome the limitations of traditional approaches toward understanding the emergence of cooperation~\cite{claus1998dynamics, kapetanakis2002coordination, McKee2023-hk,lupu2020gifting,Leibo2017-dq,hughes2018inequity,omidshafiei2017deep,cao2018emergent,rowland2021temporal,kim2020heterogeneous,foerster2016learning}. For instance, McKeen et al. have employed MARL to explore network-rewiring strategies for intervention agents and discovered more effective intervention policies than simple ones ~\cite{McKee2023-hk}. Including McKeen’s work, however, most studies incorporating MARL use hyper-parameters or reward functions to incentivize agents towards inducing cooperation, thereby not exploring different scenarios under which cooperation does or does not arise. 

On the other hand, Lupu and Precup have extended the action space of agents, instead of employing group-level optimization, to identify the conditions that facilitate the learning of cooperation~\cite{lupu2020gifting}. They incorporated a peer rewarding mechanism in their MARL simulations, which significantly improved learning progression for cooperation. Our work further extends their approach to make agents learn peer rewarding and punishing policies through network rewiring, such as attachment and detachment, while simultaneously learning cooperation.


\section{Methods}

\begin{figure*}[t]
  \centering
  \includegraphics[width=0.8\linewidth]{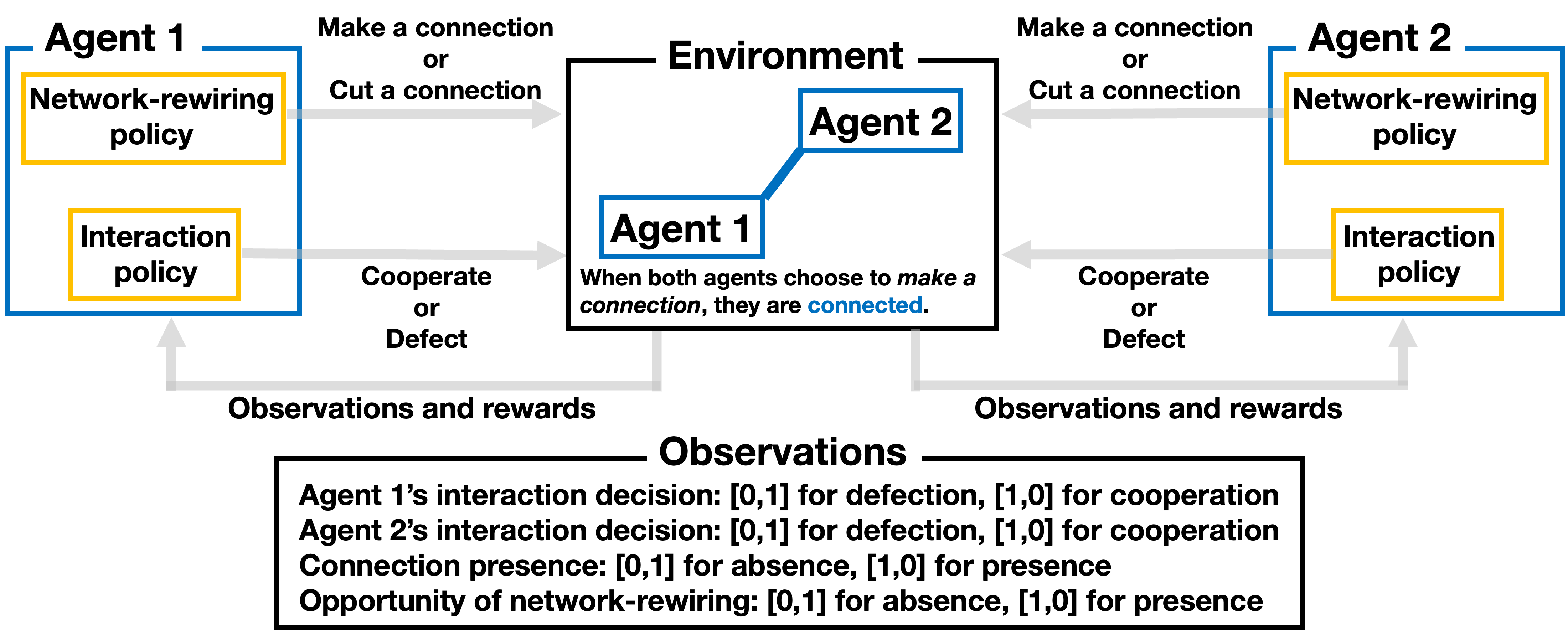}
  \caption{The overview of our simulation setting. Each agent has two sets of Q-networks: one for the interaction policy and the other for the network-rewiring policy. In the environment, two agents are connected when both choose to make a connection in a given timestep (i.e., bilateral tie-making), and are otherwise disconnected (i.e., unilateral tie-breaking). Agents play a round of Prisoner's Dilemma in a given timestep if they are connected and receive payoffs accordingly (and otherwise receive a payoff of 0 if no connection exists). Network-rewiring policy first affects the environment by making or breaking a connection before the interaction policy takes effect, thus making it possible for one agent to prevent even the possibility of cooperation at each timestep.}
  \label{fig:method}
  \Description{Method overview.}
\end{figure*}

\label{marl}
We conduct our studies in a MARL environment wherein a pair of agents simulate network-rewiring interactions across multiple episodes (Figure \ref{fig:method}). Each agent decides their behavior based on their observations, intending to maximize their respective rewards. To examine how network rewiring affects the spontaneous development of cooperation, we do not incorporate any incentive mechanisms that maximize group benefits. We perform the simulations using Acme, a framework for distributed reinforcement learning~\cite{hoffman2020acme}. We next formulate our approach in detail.\footnote{We will open source our codebase upon acceptance.}

\subsection{Formulation}
\label{formulation}
\subsubsection{Policies and action space}
\label{action_space}
Each agent in our framework makes use of two policies to make decisions at each timestep: an \textit{interaction policy} and a \textit{network-rewiring policy} (see Figure \ref{fig:method}).

Network-rewiring policy: at each timestep of a given episode, agents may have an opportunity to modify their connections to one another.
Given such a network-rewiring opportunity, each agent can choose to establish or sever connections with the other agent. 
We conduct ablations over the number of network-rewiring opportunities in a given episode (e.g., agents in some experiments are allowed to rewire at every timestep, and in others only allowed to rewire at intermittent timesteps; see Section \ref{tie_rewiring_option} for details). 
When both agents decide to establish a connection at the same timestep, the connection is established (or maintained if already present; i.e., bilateral tie-making). Otherwise, the connection is not established (or removed if already present; i.e., unilateral tie-breaking).

Interaction policy: depending on the connection status of the agents, they may then engage in a game with one another.
If agents are disconnected, no such interaction occurs between them. 
If agents are connected, then for that timestep they can each choose to cooperate or defect in a Prisoner's Dilemma scenario as described in Table\ref{tab:payoff}.
This interaction structure creates a social dilemma that favors defection over cooperation: if agents simply learn a short-sighted, greedy policy, the dyad archives the equilibrium of mutual defection that gives both no payoffs. It is, however, a worse outcome than rewarding mutual cooperation.

Moreover, to earn the cooperation benefit, it is necessary for agents to be connected. By establishing connections, however, agents are at risk of being exploited by defectors. Thus, the challenge with this interaction setting is how agents can learn to overcome myopia and exploitation risks so as to obtain public goods through network connections and mutual cooperation.

\begin{table}
  \caption{Payoff structure of the Prisoner’s Dilemma game. Each cell indicates Agent 1's payoff (left) and Agent 2's payoff (right) with indicated interaction options.}
  \label{tab:payoff}
  \begin{tabular}{ll|cc}
    \toprule
    &&\multicolumn{2}{c}{\textbf{Agent 2}}\\
    && Cooperation & Defection\\
    \midrule
    \multirow{2}{*}{\textbf{Agent 1}}& Cooperation & 1, 1 & -1, 2 \\
    & Defection & 2, -1 & 0, 0 \\
    \bottomrule
  \end{tabular}
\end{table}

\subsubsection{Observation space}
\label{observation_space}
At each timestep, each agent receives the following set of one-hot observations:
\begin{itemize}
    \item The agent's own previous interaction decision ([0,1] for defection, [1,0] for cooperation).
    \item The other agent's previous interaction decision ([0,1] for defection, [1,0] for cooperation).
    \item Whether a network edge was present in the previous timestep: ([0,1] for absence, [1,0] for presence). 
    \item Whether there was an opportunity to perform network rewiring in the previous timestep ([0,1] for absence, [1,0] for presence). 
\end{itemize}
For the last observation, agents always receive [0,1] in the first timestep of each episode because they do not have the previous timestep or the opportunity to perform network rewiring. See also the Supplementary Material for details regarding the observational space in the first timestep of each episode.

\subsubsection{Episode sequence}
\label{episode}
For each experiment trial, agents interact in 200,000 episodes, each consisting of 10 timesteps (i.e., 2,000,000 steps total). Each episode starts with two agents connected. In each timestep, the two policy networks of each agent provide network-rewiring and interaction actions, respectively (Figure \ref{fig:method}). If agents are given a network-rewiring opportunity (see Section \ref{tie_rewiring_option}), they first take the network-rewiring actions, after which they are either connected with or disconnected from one another. Otherwise, the connection is the same as the previous timestep. If the agents are connected, they then interact with their interaction actions (i.e., cooperation or defection) and receive payoffs based on their own and others' decisions (Table \ref{tab:payoff}). However, if they are disconnected, interaction does not occur, and agents receive 0 payoff for that timestep. The process is repeated until the end of the episode.

\subsubsection{Learning algorithm}
\label{learning}
For training, agents use the Double Deep Q-Network (DQN) algorithm~\cite{vanhasselt2015deep} along with prioritized experience replay~\cite{schaul2016prioritized}. Each agent has two sets of Q-networks: one for the interaction policy and the other for the network-rewiring policy. The agents' policies train independently from one another, and no parameters are shared between the agents. Due to this setting, agents cannot directly copy the other's policy, even if the other agent performs better. For each Q-network, we use a Multi-Layer Perceptron (MLP) neural network architecture with 16, 16, 2 hidden nodes in each respective layer, and $tanh$ activation function. See the Supplementary Material for more information about hyperparameters.

\subsection{Treatment ablations}
\label{plan}
We manipulate two dimensions in this study: 1) the frequency of network-rewiring opportunities and 2) the interaction policy for one of the agents. In some of our experiments, by locking one agent's policy, we examine how network rewiring encourages or discourages agents from learning about cooperation and ostracism from their counterparts. 

In total, we examine 9 treatment combinations of network-rewiring frequency and policy fixation: 3 network-rewiring opportunity conditions (no-rewiring, half-rewiring, and full-rewiring; see Section~\ref{tie_rewiring_option}) cross with 3 fixed policy conditions (no-bias, ALCC-bias, and TFT-bias; see Section~\ref{fixed_policy}). We also explore the ostracism-bias condition with the three levels of network-rewiring frequency. We conduct 15 random seeds for each combinational condition.

\subsubsection{Network-rewiring opportunities}
\label{tie_rewiring_option}
We manipulate the frequency of network-rewiring opportunities as follows. In the "no-rewiring" condition, the agents do not have the opportunity to make network-rewiring decisions. Since they are always connected, any learning driven by network rewiring does not occur in this control condition. In the "full-rewiring" condition, on the other hand, the agents have the opportunity to make network-rewiring decisions every timestep. In addition, we explore the "half-rewiring" condition where the agents make network-rewiring decisions every even timestep (i.e., 2, 4, 6, 8, and 10th timesteps) while they make interaction decisions every timestep. Thus, the agents are expected to consider a longer-term consequence of network-rewiring decisions in the half-rewiring condition. 

\subsubsection{Fixed policies}
\label{fixed_policy}
Independent of the network-rewiring opportunities, we also conduct ablations controlling the policy of one of the agents for \emph{interaction} actions. In the "no-bias" condition, no fixed policy is implemented (i.e., both agents learn their interaction policy). In the "ALLC-bias" condition, one agent has a fixed policy of always cooperating (ALLC) with the other agent. In the "TFT-bias" condition, one agent is forced to follow the Tit-for-Tat (TFT) policy, which starts with cooperation and then copies the other agent's previous interaction action. In all the conditions, both agents (including the one whose interaction policy is fixed) learn the network-rewiring policy using RL.

We design the fixed policy treatment, following the system-biology approach to explore evolutionary stable strategies ~\cite{Nowak1993-jy, Nowak2006-bl}. With policy fixation, we examine how cooperation would evolve from the seeds of a cooperative policy and how network rewiring would affect the process. It is widely accepted that ALLC agents are vulnerable to exploitation by defectors and that their cooperative policy rarely generates mutual cooperation in a standard social dilemma setting. On the other hand, TFT agents, although more complicated, could trigger the development of mutual cooperation when their counterparts seek cooperation ~\cite{Axelrod1981-xy}. TFT agents are also robust to defections as they prevent themselves from being exploited by copying defection.

In addition, we examine a condition in which we control the policy of one of the agents for \emph{network-rewiring} actions. In the "ostracism-bias" condition, one agent chooses to establish connections when the other cooperates and to serve connections when the other defected in the previous timestep. This additional condition enables us to extend our main findings by examining whether ostracism alone can develop cooperation through RL.

\subsection{Evaluation}
\label{evaluation}
For each of the treatment ablations above, we evaluate how well agents cooperate with each other. 
To measure the level of mutual cooperation, we calculate the average number of times the two agents chose to cooperate simultaneously per episode, based on the interaction structure shown in Table \ref{tab:payoff}. We only count mutual cooperation actions when the agents have a connection at a given time. For instance, if both agents are connected and cooperate five times in a 10-timestep episode, the rate of mutual cooperation for that episode is 0.5. See the Supplementary Information for further details.


\begin{figure*}[h]
  \centering
  \includegraphics[width=0.65\linewidth]{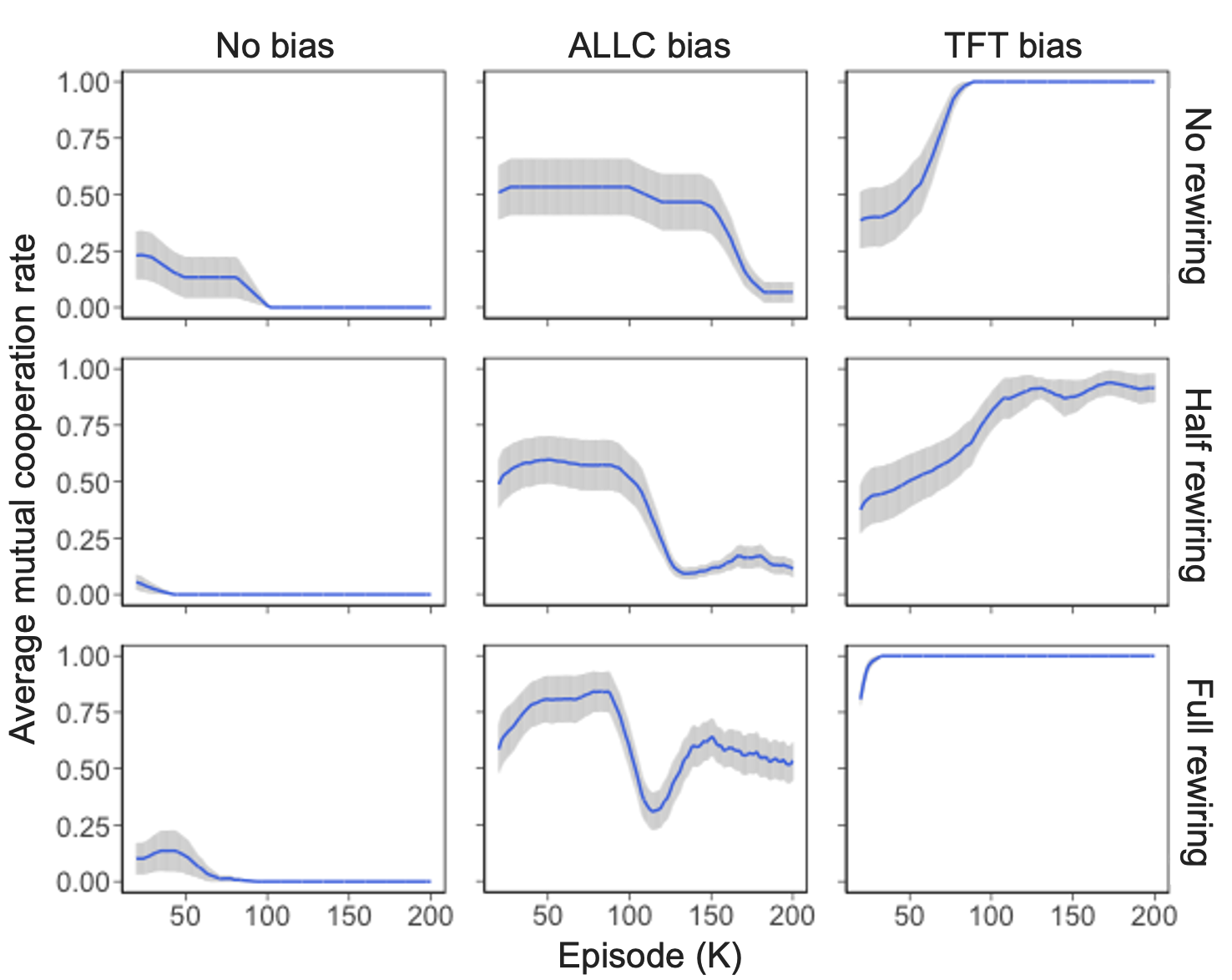}
  \caption{Average mutual cooperation rates across behavioral biases and the opportunity of network rewiring. Blue lines indicate the average rates of mutual cooperation. Gray shades indicate standard errors (\emph{N}=15). The result shows the opportunity of network rewiring facilitates the emergence of mutual cooperation when one of the agents in a group is equipped with any of the behavioral biases, and the ALLC-bias condition was more heavily affected by the amount of rewiring opportunities. 
}
  \label{fig:eval_mutual_cooperation}
  \Description{Mutual cooperation rates in the evaluation phase.}
\end{figure*}

\section{Results}
This section presents the results of our experiments, including treatment ablations.
\subsection{Effects of network rewiring on cooperation}
Figure \ref{fig:eval_mutual_cooperation} shows average mutual cooperation rates across the level of network-rewiring capability and fixed interaction policies throughout training. The comparison across the fixed policies shows that cooperation emerges with network rewiring, especially when one agent adopts a cooperative policy, such as ALLC and TFT. In keeping with prior work ~\cite{Perc2010-ac, Rand2011-hv, Shirado2013-nl}, cooperation is unlikely to arise or persist without network rewiring, except when one agent implements the TFT policy. Moreover, even with network rewiring, cooperation does not emerge among agents who learn policies from scratch (i.e., the no-bias condition) in our simulations. This result highlights the challenge of resolving social dilemmas, especially without at least some individuals who are willing to cooperate.

We next analyze the ALLC-bias column of Figure \ref{fig:eval_mutual_cooperation}. We find that enabling rewiring opportunities leads to a more frequent emergence of mutual cooperation, especially when one agent consistently chooses cooperation, i.e., the ALLC policy. In the ALLC-bias condition, cooperation collapses in the no-rewiring and half-rewiring conditions due to the other (unlocked-policy) agent taking advantage of the cooperation benefit. Simply having agents implement a generous, cooperative policy is not enough to overcome social dilemmas as free riders easily exploit it. However, if agents are given sufficient chances to rewire their connections, cooperation becomes more resilient to defection (as shown at the "ALLC bias" crossed "Full rewiring" in Figure \ref{fig:eval_mutual_cooperation}). Even in the full-rewiring condition, cooperation levels actually decrease after around 100K episodes with the ALLC-bias agents, but then bounce back. This result suggests that network rewiring enhances cooperation resilience through reinforcement learning, as cooperators can defend themselves against defectors by detachment.

We next discuss the TFT-bias column of Figure \ref{fig:eval_mutual_cooperation}. When an agent implements the TFT policy, which involves starting with cooperation and following their counterpart's interaction, most random seeds result in mutual cooperation across the rewiring conditions. This suggests that mutual cooperation relies less on learning the network-rewiring policy in the TFT-bias condition than in the ALLC-bias condition. The TFT policy allows the agents to prevent exploitation from free riders by copying defection, which means that further opportunities to detach from defectors may be less impactful for the group dynamics. 

Nevertheless, network rewiring can accelerate the process of mutual cooperation in conjunction with the TFT policy. In the TFT-bias full-rewiring condition, agents' interaction actions quickly converge to mutual cooperation and never collapse afterward. It is worth noting that the development of mutual cooperation is slower in the half-rewiring condition than in the no-rewiring condition, possibly due to fewer learning opportunities for their interaction policies. If agents initially sever connections, they earn zero rewards regardless of their interaction actions (i.e., cooperation and defection), which suppresses their learning about how they should interact with each other. Network rewiring affects the developmental process of cooperation by balancing reciprocal opportunities for connections and learning opportunities for cooperation.

\begin{figure}[h]
  \centering
  \includegraphics[width=0.9\linewidth]{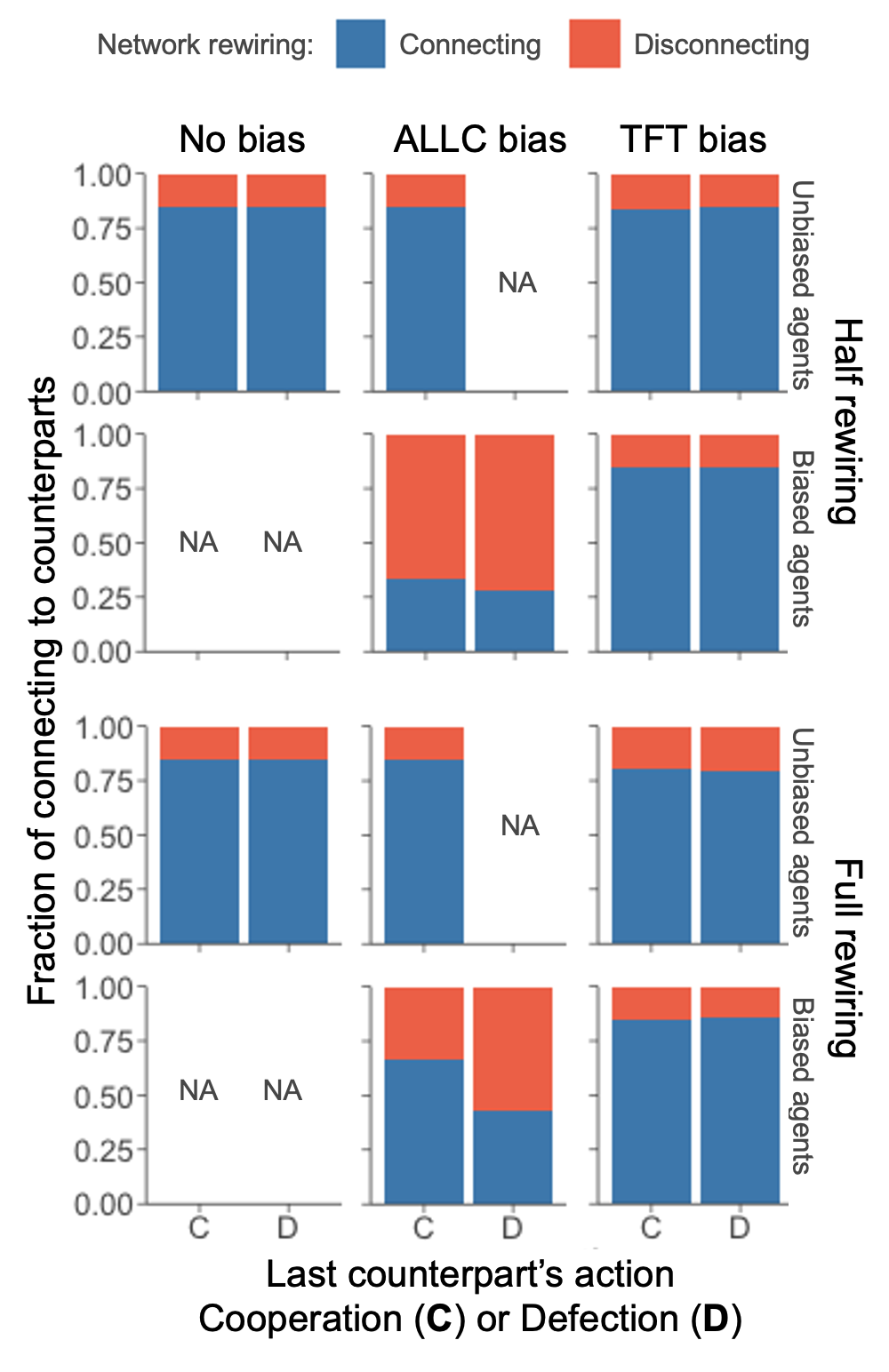}
  \caption{Fraction of agents establishing and severing connections across the conditions after reinforcement learning of cooperation and network rewiring. Blue bars indicate the fractions that agents \emph{intend} to connect to their counterparts, instead of disconnecting from them (red bars). When both agents choose to connect at the same timestep, the connection is established or maintained. Otherwise, the connection is not established or removed.  ALLC-biased agents exhibit higher rates of ostracism: attempting to establish connections with the other agent that cooperated and severing connections with the other that defected in the last timestep. }
  \label{fig:rewiring_matrices}
  \Description{Stacked bar graphs that describe the rewiring response in each condition}
\end{figure}

\subsection{Learned network-rewiring policies}
We then turn to how agents develop their network-rewiring policies through interactions with the social dilemma. Figure \ref{fig:rewiring_matrices} shows the fractions of connecting to the counterparts (instead of disconnection from them) based on the data obtained from the end phase of training. These results indicate, after substantial interactions and learning, whether an agent establishes or seizes the connection with their counterpart based on their interaction action. 
Note that timesteps where agents do not have the opportunity to rewire in the half-rewiring condition are excluded from the analysis.
For instance, the top-left cell illustrates the results of the no-bias half-rewiring condition. In this condition, agents attempt to connect or maintain a connection with their counterparts 84.6\% of the time (instead of disconnecting from them) when their counterparts have cooperated in the last timestep and 85.3\% of the time when they have defected. 

Our analysis shows that agents generally prefer to establish connections rather than sever them. Regardless of whether their counterparts cooperate or defect, agents in the no-bias and TFT-bias conditions choose to have connections with their counterparts more than 80\% of the time. This is an interesting outcome, indicating that agents prefer to connect with each other to enable interactions and the possibility of earning payoffs.

However,  the network-rewiring policy correlates with cooperation in the opposite way between the no-bias and TFT-bias conditions. Choosing to connect associates with mutual defection in the no-bias condition, where most agents start with defection (Figure \ref{fig:eval_mutual_cooperation}). Defectors face no risk of exploitation, so they seek out interactions for possible earnings, which makes them prefer connecting. It makes it more difficult for agents to change their behavior to cooperation. In contrast, in the TFT-bias condition, agents with the TFT policy can manage the risk of exploitation by choosing defection when facing defectors. This increases the expected benefit of connection even when their counterparts have chosen defection in the last timestep. As mutual cooperation evolves with the TFT policy (Figure \ref{fig:eval_mutual_cooperation}), the advantage of having connections increases.

In the ALLC-bias scenario, agents develop a more complex policy for network rewiring (Figure \ref{fig:rewiring_matrices}). Agents with always cooperation learn to ostracize other agents in the full-rewiring condition. They connect with cooperators 63.4\% of the time, while they connect with defectors only 39.9\% of the time. This shows that agents eventually value the quality of connections over the quantity during network rewiring. In fact, the learned rewiring bias is similar to how humans behave in similar rewiring opportunities in a cooperation scenario~\cite{Fehl2011-kq, Shirado2013-nl}. 

On the other hand, when agents have half the network-rewiring opportunities, they do not learn to connect with cooperators while they learn to disconnect from defectors. They connect with cooperators only 32.7\% of the time, while they connect with defectors 28.5\% of the time in the half-rewiring condition. This finding suggests that agents need sufficient rewiring opportunities \textit{vis-a-vis} interaction ones to develop selective rewiring policies based on the other's past actions. 

Finally, the ALLC-bias condition does not allow unbiased agents to develop a network-rewiring policy with defectors because their counterparts always choose cooperation, regardless of network-rewiring opportunities.

Our findings suggest that, given sufficient network-rewiring opportunities, unconditional cooperators can learn the policy of ostracism, which leads to resilient cooperation (as shown at the "ALLC bias" crossed "Full rewiring" in Figure \ref{fig:eval_mutual_cooperation}). In contrast, conditional cooperators, such as TFT agents, rarely learn ostracism (as shown in the "TFT bias" column of Figure \ref{fig:rewiring_matrices}) because the benefit of connection outweighs its cost. The stable learning of ostracism depends on the unconditional cooperation of agents. Once ostracism is established, it prevents cooperation from dissolution.

\begin{figure}[h]
  \centering
  \includegraphics[width=1.0\linewidth]{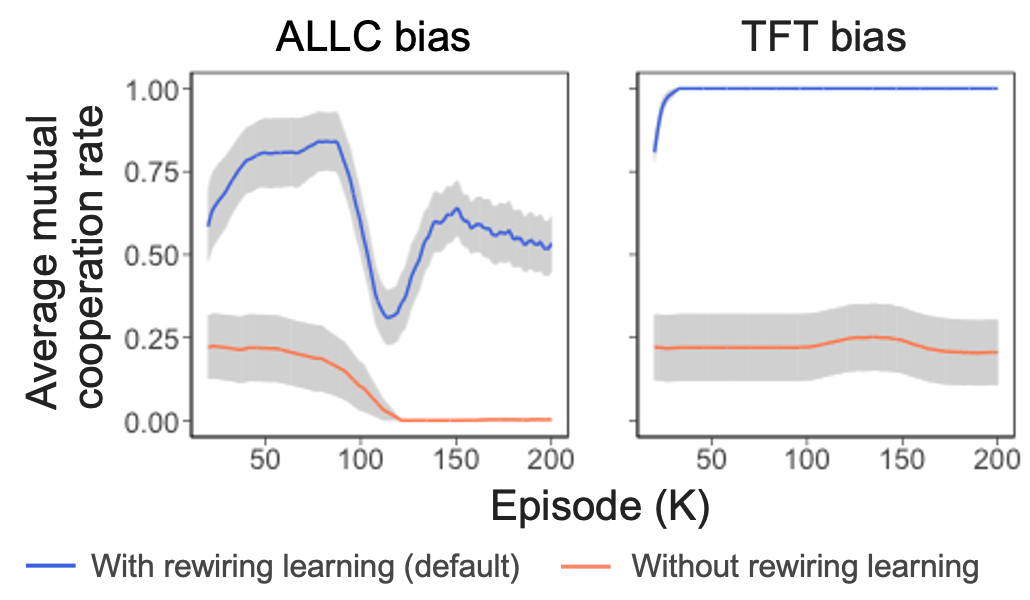}
  \caption{Average mutual cooperation rates in the ALLC-bias and the TFT-bias when the learning of rewiring action is disabled (red lines) and enabled (blue lines). All are in the full-rewiring condition. Lines indicate the average rates. Shades indicate standard errors (\emph{N}=15). We show the results  "with rewiring learning" for comparison, which are identical to those in the full-rewiring condition shown in Figure \ref{fig:eval_mutual_cooperation}.}
  \label{fig:random_rewiring_mutual_cooperation}
  \Description{Mutual cooperation rates in random rewiring conditions}
\end{figure}

\subsection{The \emph{learning} of ostracism, rather than the opportunity of rewiring, facilitates cooperation}

Next, we deconstruct and pinpoint the causal effects of network rewiring on cooperation. To do so, we disable the learning of the network-rewiring policy, isolating the effects of rewiring opportunities \emph{per se}. We set the learning rate of the network-rewiring policy to zero for both agents in each group, allowing the agents to exhibit network-rewiring behavior as in the initial random policy. We only permit the agents to update the interaction policy governing whether to cooperate.

As shown in Figure \ref{fig:random_rewiring_mutual_cooperation}, when the updates of the neural network for the network-rewiring policy are disabled, the cooperation level is significantly lower in both the ALLC-bias and TFT-bias conditions compared to when the learning is allowed (which is identical to the result of the full-rewiring condition in Figure \ref{fig:eval_mutual_cooperation}). Agents develop almost no mutual cooperation without rewiring-policy learning, which is more defective than the interactions without rewiring opportunities (i.e., the no-rewiring condition). 

This is because, without learning, agents randomly choose to connect and disconnect, and consequently, they mutually establish connections and interact only 25\% of the time on average. Thus, the maximum level of mutual cooperation is limited to 25\%. Furthermore, network-rewiring opportunities solely reduce the opportunities to learn their interaction policies by intermittent disconnections. This is salient in both bias conditions. In the ALLC-bias condition, agents resume mutual cooperation after a halfway collapse with network-rewiring learning, but they do not without learning. In the TFT-bias condition, the level of mutual cooperation rises to 100\% in the first 30K episodes with network-rewiring learning, while no such increase occurs without learning. 

The development and sustainability of cooperation through network rewiring is due to agents developing a specific network-rewiring policy, such as ostracism, through learning from interactions. In other words, if agents lack the ability to learn an effective way to rewire connections, they will not achieve mutual cooperation even with network rewiring. 
Thus, the learning of ostracism, rather than the opportunity for rewiring, matters in fostering cooperation.

\begin{figure}[h]
  \centering
  \includegraphics[width=0.6\linewidth]{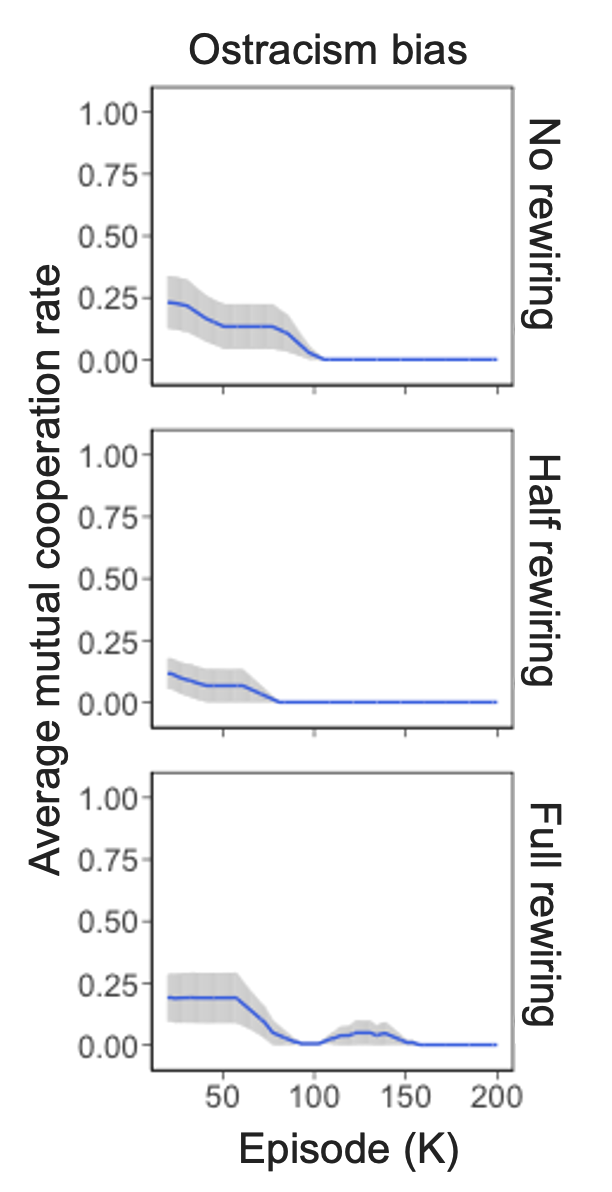}
  \caption{Average mutual cooperation rates in the ostracism-bias condition across the rewiring conditions. In the ostracism-bias condition, one of two agents in a group always chooses to establish connections when the other cooperates and to serve connections when the other defected in the previous timestep. This result suggests that cooperation rarely arises from ostracism alone.}
  \label{fig:ostracism_mutual_cooperation}
  \Description{Mutual cooperation rates in the Ostracism-bias condition}
\end{figure}

\subsection{Cooperation does not emerge from ostracism alone}
Our findings suggest that learning network-rewiring policies like ostracism is critical in addressing social dilemmas. Acquiring the ability to cooperate first as a behavioral predisposition facilitates the acquisition of ostracism behaviors, leading to the emergence of mutual cooperation. However, it remains unclear whether ostracism can foster cooperation in the absence of conditional or unconditional cooperators. To address this, we conduct additional simulations in the "ostracism-bias" condition, where one agent always aims to connect with the other that cooperated in the previous timestep, and disconnects from those that defected earlier.

We find that in the ostracism-bias condition, agents did not establish mutual cooperation in any of the rewiring conditions (Figure \ref{fig:ostracism_mutual_cooperation}). The agents with ostracism abilities can defect just like the no-bias agents. Thus, as in the no-bias condition, the interaction quickly leads to mutual defection in the ostracism-bias condition, providing no opportunity for agents to experience selective network rewiring through ostracism. These findings suggest that \emph{extrinsically} endowing agents with the capacity to ostracize others does not enable mutual cooperation. Instead, as shown in the previous section, it is crucial for agents to learn ostracism behaviors \emph{spontaneously} through cooperative bias to develop and sustain mutual cooperation.

\section{Discussion}
Our study shows that the ability to rewire connections for interaction is highly correlated with the emergence of mutual cooperation. When agents have more opportunities to establish and sever their connections, they achieve mutual cooperation more often. This finding is in keeping with prior human studies showing that cooperation emerges with network rewiring~\cite{Rand2011-hv, McNamara2008-io, Shirado2013-nl, McKee2023-hk}. By using a two-policy approach and MARL, our study further clarifies that the network-rewiring effect is exerted in the presence of behavioral biases~\cite{Nowak1993-jy, Nowak2006-bl}. In our simulations, only when one agent implements a simple cooperative policy (ALLC) or complex reciprocal one (TFT), can agents have mutual cooperation with rewiring capabilities. Building on such cooperative agents, our study shows that network rewiring can make cooperation resilient against defection.

We find that network rewiring has both advantages and disadvantages in the evolution of cooperation. As prior work suggested, enabling network rewiring can help cooperators prevent defectors from taking advantage of them~\cite{Rand2011-hv, Shirado2013-nl}. On the other hand, our study shows that it can also limit the opportunities for agents to interact with each other, earn payoffs, and learn their policies. For example, network rewiring delays the development of cooperation in the half-rewiring TFT-bias condition, suggesting the disadvantages outweigh the advantages at lower connection mobility. Our findings suggest that for agents to use the benefits of rewiring opportunities for cooperation, they might need to learn a selective rewiring policy, such as ostracism, through limited interaction opportunities. It could also be favorable for cooperation that a dynamic network consists of diverse individuals with biased policies, including unconditional or conditional cooperators~\cite{Santos2008-qf}.  

We clarify the role of ostracism in underlying cooperation. In our study, when network-rewiring opportunities are present, cooperation-biased agents are more effective at learning ostracism than agents without a behavioral bias or with the TFT policy. While no-bias and TFT agents prefer connecting to their counterparts, regardless of their interaction actions, always-cooperating (ALLC) agents choose whether to connect based on whether their counterparts cooperate or not. This finding suggests that unconditional cooperators or "zealots" might be necessary to establish ostracism, which facilitates group cooperation~\cite{Masuda2012-qc, Shirado2020-hs}. We also confirm that cooperation rarely emerges from ostracism alone. Rather, ostracism can act as a catalyst that helps dynamic networks favor cooperation over defection.

There are other features potentially relevant to the coevolution of cooperation and ostracism that we have not explored in this study. For example, we do not observe mutual cooperation in the no-bias condition, even with the opportunity for network rewiring. However, our findings also suggest that mutual cooperation would be possible if at least one agent learned, for example, the TFT policy. It is an important next step to address how agents can learn such a complex, reciprocal policy through interaction and network rewiring in the setting we considered. We find that agents can learn ostracism, which is a reciprocal policy for network rewiring, from a simple cooperation bias. Thus, the TFT might also be developed from simple behavioral biases. 

Additionally, our study only considers symmetric interactions between two agents, as this setting permits a more scrutinous analysis of inter-agent interactions. Nonetheless, expanding the setting to larger multi-agent systems with complex network structures and dynamics could provide further insights into the coevolution of cooperation and ostracism. Moreover, other interaction strictures, such as asymmetric cooperation benefit and cost between agents ~\cite{Janssen2011-mw, Nishi2015-uh} or explicit costs of network connections~\cite{Marwell1988-kv}, might drive reinforcement learning and interaction to different equilibria with network rewiring. Finally, the learning process in our simulations is guided by an off-policy algorithm, the Double Deep Q-Network (DQN). Thus, agents learn their policies partially independently from their latest observations. Future studies could compare the potential differences between off-policy and on-policy algorithms, such as Proximal Policy Optimization (PPO)~\citep{schulman2017proximal}, to further understand how learning processes can influence the emergence of cooperation and ostracism.

In conclusion, our study highlights the importance of network rewiring and the learning of ostracism in the coevolution of cooperation and ostracism in MARL settings. Network-rewiring opportunities could help a few individuals having a cooperative tendency learn ostracism, which facilitates, in turn, cooperation over a population. These findings contribute to understanding how and when different types of policies, such as interaction and network-rewiring, could emerge simultaneously in multi-agent systems with network dynamics.






\bibliographystyle{ACM-Reference-Format} 
\bibliography{sample}


\end{document}